\documentstyle[prd,epsf,aps,preprint]{revtex}
\tighten

\newcommand{\lsim}{\lower .5ex\hbox{$\buildrel < \over {\sim}$}}

\preprint{\vbox{Submitted to Physical Review Letters \hfill }}  

\begin{document}


\title{Many--Body Corrections to Charged--Current Neutrino Absorption Rates
in Nuclear Matter}
\author{Adam Burrows}
\address{Department of Astronomy, The University of Arizona, Tucson, AZ 85721
\\ e-mail: burrows@as.arizona.edu}
\author{R. F. Sawyer}
\address{Department of Physics,
The University of California,
Santa Barbara, CA 93106
\\ e-mail: sawyer@sarek.physics.ucsb.edu}
\maketitle
\setcounter{page}{1}
\begin{abstract}

Including nucleon--nucleon correlations due to
both Fermi statistics and nuclear forces, we have
developed a general formalism for calculating
the charged--current neutrino--nucleon absorption rates in
nuclear matter.  We find that at one half nuclear
density many--body effects alone suppress the rates by
a factor of two and that the suppression factors increase to
$\sim$5 at $4\times10^{14}$ g cm$^{-3}$.  The associated
increase in the neutrino--matter mean--free--paths
parallels that found for neutral--current interactions and
opens up interesting possibilities in the context
of the delayed supernova mechanism and protoneutron star cooling.

\end{abstract}




\section{Introduction}

The neutrino absorption and scattering opacities in the post--shock
core of a supernova, in which nuclei are largely disintegrated into nucleons,
determine the duration, spectrum, and flavor distribution of the emerging
neutrino pulse. It has been known for some time that the interactions
among nucleons in the denser regions can change these opacities
significantly, but to date there has been no comprehensive treatment given in the
literature and present calculations of the complete supernova process do not
include the effects of interactions on the opacities. 

The neutrino--matter interaction rates can be related to the space-- and time--dependent
correlations among the set of density operators for the separate nuclear
constituents (to find the Gamow--Teller parts we must consider separate
spin--up and spin--down densities). In the case of neutral--current
interactions \cite{BS}, there is an instructive limit, which also provides an estimate
of the effects, in which the combined limits of large nucleon mass and
small neutrino energy allow the use of long--wavelength limits of equal--time
correlation functions, in turn expressible in terms of the second
derivatives of an energy density functional with respect to various
densities. This approach is the direct multichannel generalization of the
familiar results for light scattering from the thermal density fluctuations
in a fluid, where it is the compressibility that determines the 
long--wavelength opacity, and it was used in references \cite{s2Ray} and
\cite{iwa}
to find significant reductions of neutral--current opacity in certain regions.
In Burrows \& Sawyer \cite{BS}, an approach based on ring graphs was used to
encompass these results and to extend them to domains in which the
equal--time and long--wavelength limits are not clearly applicable.

The use of the equal--time and long--wavelength limits to express correlation functions in
terms of static susceptibilities cannot be extended to the charged--current
interactions when there is a large chemical potential difference between protons
and neutrons. Furthermore, there do not exist in the present
literature systematic estimates of the effects of interactions
on the charged--current opacities for electron neutrinos. In the present work, 
we give a theoretical
framework for addressing these opacities, based on summing ring graphs,
together with the results of calculations with input parameters taken from
the current phenomenology of nuclear matter.

\section{ Many--Body Formalism for Charged--Current Rates}

The charged--current interactions of the nuclear medium are determined by
the statistical averages involving the nucleonic charged--current operator,
$j_{\mu}=\bar\psi_p\gamma_{\mu}(1-g_A\gamma_5)\psi_n$,  and its
Hermitean conjugate,
\begin{equation}
W_{\mu\nu}(q,\omega)=-iZ^{-1}\int d^4x \;e^{-i \bf q
\cdot \bf x}\rm\it e^{i(\omega+\hat\mu) t}Tr\bigl[e^{-\beta( H-\Sigma
\mu_i N_i)}[j_{\mu}(\bf x \rm\it , t),j^{\dagger}_{\nu}(0,0)]\bigr]\theta
(t),
\label{1}
\end{equation}
where $Z$ is the partition function and $\hat\mu$ is the chemical
potential difference, $\hat\mu=\mu_n-\mu_p$. The vector, $({\bf q},\omega)$, is the
momentum--energy transfer from the leptons to the medium. The transition rate,
differential in angle and $\omega$,  for the reaction $\nu_e + n \rightarrow e^{-} + p$ 
is given in terms of $W_{\mu\nu}$ as 
\begin{equation}
\frac{d^2 \Gamma(\nu_e \rightarrow e^-)}{d \omega \hspace{1 pt} d
\cos(\theta)}=(2 \pi^2)^{-1}G_W^2
\frac{(E_1-\omega)^2[1-f_{e}(E_1-\omega)]}{1-\exp[-\beta(\omega+
\hat\mu)]}\Lambda^{\mu\nu}(q,\omega) {\rm Im}W_{\mu\nu}(q,\omega),
\label{2}
\end{equation}
where 
\begin{equation}
\Lambda^{\mu\nu}=(4E_1E_2)^{-1}Tr[p_1 \hspace{-10pt}/ \hspace{8
pt}(1-\gamma_5)
\gamma^{\mu}p_2 \hspace{-10
pt}/\hspace{8pt}\gamma^{\nu}(1-\gamma_5)] \, ,
\label{3}
\end{equation}
$q=[E_1^2+(E_1-\omega)^2-2E_1(E_1-\omega)\cos\theta]^{1/2}$,
and $E_1$ in the incident neutrino energy.
To verify the factors in (\ref{2}) replace the
commutator in (\ref{1}) by the unordered product, the $\theta(t)$
function by 1/2 in order to capture the imaginary part, and 
the statistical factor in the denominator by unity. Then, the introduction of
a complete set of states between the current operators gives the inclusive
$\nu_e\rightarrow e^-$ rate. Recalling that the Heisenberg picture for the
density operators is defined with respect to the operator $ H-\Sigma
\mu_i N_i $, we note that the $e^{i\hat\mu t}$ factor in (\ref{1}) will be
canceled
before the time integral is performed.  The other term in the commutator
gives the same answer, but with an additional factor,
$-\exp[-\beta(\omega+\hat\mu)]$. The statistical factor in the denominator
of (\ref{2})  cancels when the two terms are combined. 

Taking the nucleons as nonrelativistic, which
is allowable up to about twice nuclear density, we note that the components
$W_{i0}$ and $W_{0i}$ vanish, and that the space tensor, $W_{ij}$,
derives from the axial--vector current and can be written in the form, 
\begin{equation}
W_{ij}(\omega,\cos \theta) =g_A^2W_A (\omega,\cos
\theta)\delta_{ij}+g_A^2 W_T(\omega,\cos \theta)q_i q_j\, .
\label{4}
\end{equation}
We define $W_{00}= W_V$, as it comes entirely from the vector
current. Combining with (\ref{3}) and calculating the lepton trace, we
now have
\begin{equation}
W_{\mu\nu}\Lambda^{\mu\nu}=2(1+\cos
\theta)W_{V}+2(3-\cos\theta)g_A^2 W_A+2\omega^2(1+\cos
\theta)g_A^2 W_T .
\label{5}
\end{equation}
We calculate the three functions $W_V$, $W_A$, and $W_T$ in the ring
approximation. For these graphs, a unit of charge is passed in each
nuclear interaction in the expansion of the charged--current correlation
function of (\ref{1}). If the potential between nucleons $a$ and $b$
conserves isotopic spin, this means that only the terms containing the
isotopic operator $\vec \tau^{(a)}\cdot\vec\tau^{(b)}$  enter. We assume a
potential in momentum space,
\begin{equation}
V^{a,b}(q,\omega)= \bf\vec\tau \it^{ a} \cdot\vec\tau^b [
v_1(q)+\vec\sigma^a \cdot\vec\sigma^b \it v_2(q)+\vec\sigma^a \cdot
\vec{q}\vec\sigma^b\cdot\vec{q} v_3(q,\omega)]\, .
\label{6}
\end{equation}

We have inserted in the $v_3$ term a dependence on energy transfer, $\omega$,
in order to accomodate the one--pion exchange force,
\begin{equation}
v_3(q,\omega)=-f^2 m_{\pi}^{-2}(q^2+ m_{\pi}^{2}-\omega^2)^{-1} ,
\end{equation}
with $f^2 \approx 1$. 

For the other two potentials, we follow the development
of \cite{BS} in taking zero--range forms fitted to the Landau parameters of
reference \cite{s5Ray} and obtain
$v_2=3.4m/m^* \times 10^{-5} ({\rm MeV})^{-2}$ and $v_1=
1.88m/m^* \times 10^{-5} ({\rm MeV})^{-2}$,
where $m^*$ is the effective nucleon mass in the medium.  We
are assuming isospin invariant forms, despite the fact that our application
will be to unsymmetric matter. This appears to us to be the state of the
art. It is not totally satisfactory, in view of the fact that the $v_1$ and $v_2
$ terms are phenomenological forms that obtain in nuclear matter rather
than forms based on elementary meson exchanges. Note that in the case
of non--symmetric matter isospin symmetry is still broken through the
polarization  functions. We define the nucleon--charge--raising polarization part,
$\Pi(q,\omega)$, in parallel with (\ref{1}), in terms of the retarded
commutator of the density $n_c=\psi^{\dagger}_p \psi_n$ and its
Hermitean conjugate,

\begin{equation}
\Pi(q,\omega)=-iZ^{-1}\int d^4x \;e^{-i \bf q
\cdot \bf x}\rm\it e^{i(\omega+\hat{\mu}) t}Tr\bigl[e^{-\beta( H-\Sigma \mu_i
N_i)}[n_c(\bf x \rm\it , t),n^{\dagger}_c(0,0)]\bigr]\theta (t)\, .
\label{a13}
\end{equation}
Note that the function that gives the vector part of the rate,
$W_V(q,\omega)$, is exactly given by $\Pi$. The ring approximation \cite{FW} for
$\Pi$ is now
\begin{equation}
W_V(q,\omega) =
\Pi(q,\omega)=\frac{\Pi^{(0)}(q,\omega)}{1-2v_1(q)\Pi^{(0)}(q,\omega)}\, ,
\label{8}
\end {equation}
where $\Pi^{(0)}$ is the polarization function in the absence of
interactions. The factor of 2 multiplying the potential comes from the
isospin operator in the potential (\ref{6}). The spin dependent parts of
the potential do not contribute.

For the axial contribution, the operative non--relativistic forms of the
current are the operators  $n_c^i=\psi^{\dagger}_p\sigma^i \psi_n$, and
the analogue to (\ref{a13}) will be a tensor with indices $ i,j$. In the
absence of interactions, this tensor is given by  $\delta_{i,j}\Pi^{(0)}$.
Thus, summing the axial chain to get  $W_A{(q,\omega)}$ we obtain
exactly the same structure as  (\ref{a13}) with $v_2$ replacing $v_1$,
but with the same function $\Pi^{(0)}$,
\begin{equation}
W_A(q,\omega)=\frac{\Pi^{(0)}(q,\omega)}{1-2v_2(q)\Pi^{(0)}(q,\omega)}\, .
\label{9}
\end {equation}
The potential $v_3$ does not enter (\ref{9}), even though it couples to
the axial--vector terms, since any ring graph chain in which at least one
$v_3$ participates becomes a contribution to $W_T$. Elementary
combinatorics for the tensor chain gives,
\begin{equation} 
W_T(q,\omega)=\frac{[\Pi^{(0)}(q,\omega)]^2v_3(q)}{1-2[v_2(q)+q^2v_3
(q,\omega)]\Pi^{(0)}(k,\omega)}\, .
\label{10}
\end {equation}

The requisite polarization function is given by
\begin{equation}
\Pi^{(0)}(q,\omega)=-2\int \frac{d^3p}{(2\pi)^3}\frac{f(| \bf p \it
|,\mu_n)-f(|\bf p+q \it |,\mu_p)}{\omega+\epsilon_{\bf p}-\epsilon_{\bf
p+q}+i\eta} ,
\label{11a}
\end{equation}
where the functions $f$ are the nuclear Fermi occupation functions for
the indicated momenta and chemical potentials. We conceptually
extend (\ref{11a}) to include the average potentials that the
nucleons experience in the medium, $v_{p,n}$, by making the
replacements $ \bf( p)^2\it/(2m) \rightarrow \bf (p)^2\it/(2m)+v_n$
and $ \bf( p+q)^2\it/(2m) \rightarrow \bf (p+q)^2\it/(2m)+v_p$, both in the
denominator and in the distribution functions in the numerator.  However, we
suppose that we are starting with a table of densities and temperatures ($T$) from
an equation of state that already takes into account the potentials $v_{p,n}$.
If we utilize
Fermi distributions in which the chemical potentials are derived from the
input densities using the free--particle relations,
these average potential corrections are automatically included 
and the parameters $v_{p,n}$ do not appear explicitly in the formalism.
To do the computations, we use the following form for $\Pi^{(0)}$ :
\begin{eqnarray}
\Pi^{(0)}(q,\omega)=\frac{m^2}{2\pi^2q
\beta}\Bigl(\int_{-\infty-i\epsilon}^{\infty-i \epsilon} ds s^{-1}\log 
[1+e^{-(s+Q_+)^2+\beta \mu_n} ]
\nonumber\\
+\int_{-\infty+i\epsilon}^{\infty+i \epsilon} ds s^{-1}\log 
[1+e^{-(s+Q_-)^2+\beta \mu_p} ]\Bigr) ,
\label{15}
\end{eqnarray}
where
\begin{equation}
Q_{\pm}=\bigl(\frac{m\beta}{2}\bigr)^{1/2}
\bigl(\mp\frac{\omega}{q}+\frac{q}{2m} \bigr) ,
\end{equation}
which gives
\begin{equation}
{\rm Im} \Pi^{(0)}(q,\omega)=\frac{m^2}{2\pi \beta q} \log
\Bigl[\frac{1+e^{-Q_+^2+\beta \mu_n}}{1+e^{-Q_+^2+\beta
\mu_p-\beta\omega}} \Bigr]
\label{13}
\end{equation}
and
\begin{equation}
{\rm Re} \Pi^{(0)}(q,\omega)=\frac{m^2}{2\pi^2q \beta}\int_0^{\infty}
\frac{ds}{s}\log  \Bigl[\frac{1+e^{-(s+Q_+)^2+\beta
\mu_n}}{1+e^{-(s-Q_+)^2+\beta \mu_n}} \Bigr]\; \; 
+(\omega \rightarrow -\omega, \mu_n\rightarrow\mu_p)\, .
\label{a24}
\end{equation}

The imaginary part of the polarization (\ref{13}) can be obtained by
direct integration, and is the same as that given in references \cite{BS,reddy}. Given this,
the full function (\ref{15}) can be verified by checking the analytic properties in
the $\omega$ plane.

 We look at the results of medium interactions for two sets of
conditions typical of the dense interior of a protoneutron star, but at different
times; first for an early
time ($ t < 5 $ seconds), when the lepton number is large and second at a
later time ($ t > 10$ seconds), 
after which the lepton excess has largely radiated away \cite{BL}.

\section{Lepton--Rich Era}

The densities of the various species are such that at zero temperature the
inequality $2p_F^{(p)}>|p_F^{(n)}-p_F^{(\nu)}| $ holds. This inequality
allows the single nucleon process to proceed at full strength. That is to
say, at low temperatures the function  {\rm Im}$ \Pi^{(0)}(q,\omega)$
(\ref{13})
is large in the region of $(q,\omega)$ defined by the two conditions: 1)
the neutrino energy, $E_1$, is near the neutrino Fermi surface; 2) the
electron energy, $E_1-\omega$, is near the electron Fermi surface. These
conditions come from the occupation factors that enter the expression for
the total rate. 

In this region, we use (\ref{8}), (\ref{9}), and (\ref{10}) to
calculate the modification factors in the medium. We have estimated the
contribution of the tensor term  (\ref{10}) relative to the two other terms
and conclude that it is less than 10\% of the total in the cores of protoneutron
stars and supernovae.  For the vector and axial--vector terms, 
we compute suppression factors, $S_A$ and $S_V$, defined as the ratio of the
rates calculated with the nuclear interactions to those calculated without.
This is done by substituting (\ref{8})
and (\ref{9}) into (\ref{5}), multiplying (\ref{2}) by the
neutrino occupation function, and integrating over neutrino energies. In
Table 1, we give results using a post--bounce supernova profile 
taken from Burrows, Hayes, \& Fryxell \cite{bhf}.
As seen in the table, the Gamow--Teller suppression factors are larger
than the Fermi suppression factors.  Furthermore, the degree of many--body suppression
increases with density and decreases with temperature.  Importantly, the magnitude
of the effect above $10^{14}$ g cm$^{-3}$ is large, ranging from a factor of 2 at $10^{14}$ g cm$^{-3}$
to a factor of $\sim$5 near $4\times10^{14}$ g cm$^{-3}$.  Correspondingly, the neutrino--matter absorption
cross sections decrease with density.  Since it has recently been shown \cite{BS} that
the neutral--current scattering rates at high density are also reduced,  we conclude that
post--bounce supernova cores are significantly more transparent than 
previously believed.  This enhanced transparency should translate 
at late times ($> 500 - 2500$ ms after bounce)
into higher neutrino luminosities, that thereby may be more efficient at reenergizing
or powering a stalled supernova shock \cite{BS}.

Integrating equation (\ref{2}) over $cos\theta$,  the distribution
of the energy transfer, $\omega$, to and from the nucleons
due to the process $\nu_e + n \rightarrow e^{-} + p$ in the
lepton--rich era can be derived and is depicted in Figure 1 for a variety
of densities, from $10^{13}$ g cm$^{-3}$ to $10^{15}$ g cm$^{-3}$.
For these curves, the temperature is 5 MeV,
the electron fraction, Y$_e$, is 0.26,
the incident electron neutrinos are on their Fermi surfaces, and beta
equilibrium is assumed.
The highest curve on the right (that for $\rho$ = $10^{13}$ g cm$^{-3}$)
ignores many--body effects, though it incorporates the full kinematics,
and is included for comparison.
As expected, the peak of the energy transfer
is generally near $-\hat{\mu}$ (given in the figure caption),
since the electron blocking factor in (\ref{2})
puts the electrons on the electron Fermi surface and beta equilibrium
requires that $\mu_e = \hat{\mu} + \mu_{\nu_e}$.  (Note that $\hat{\mu}$ increases
with density.) There is a modest spread in
$\omega$ around the peak with approximately a gaussian distribution.  The width
of this distribution scales with the temperature.   Figure 1 demonstrates
what Table 1 also reflects that the total cross sections, the integrals under the
unnormalized curves, are decreasing functions
of density.

\section{Lepton--Poor Era}

As discussed in \cite{soni} and \cite{haensel}, as the trapped electron lepton number
decays, we reach a configuration in which the neutrino absorption process
discussed above dies almost completely for low temperatures. At the end of
deleptonization, we have $p_F^{(\nu)}=0$~\cite{BL}. The neutrinos then have thermal
energies and we find that the proton fraction has decreased to the point that
$2p_F^{(p)}<<p_F^{(n)}$. In this case we cannot conserve momentum for the
three degenerate species, $e^-, p , n $, when the momentum of the neutrino is
small and when we stay near the Fermi surface for the three other species. At
low temperatures, the function Im$ \Pi^{(0)}(q,\omega)$ of (\ref{13}) is now
exponentially small in the region of $(q,\omega)$ defined by the leptonic
occupancies. Thus, the ring graphs, as defined in the previous section, give a 
negligible rate for the charged--current processes at low lepton number and temperature.

The other mechanisms that have been used to estimate the rates 
depend on a spectator nucleon to transfer the necessary momentum,
either through a potential or through an assumed correlation \cite{soni,haensel,FM}.
Translated to graphs, these mechanisms involve the estimation
of proper graphs for the polarization parts, where a
proper graph is defined as a graph that cannot be cut into two 
disjoint parts by severing a single potential line. However, there 
are ``ring corrections" to such graphs, in which the initial or final
current vertices attach to a ring chain that then attaches to the
proper polarization graphs. These ring corrections then have a big
suppressive effect on the primary mechanism for momentum transfer,
as one can see from the following argument.

We consider a proper polarization graph in which a nucleon--nucleon
interaction has intervened to allow momentum conservation near the Fermi
surfaces. That is to say, by (\ref{9}), we have a term for the vector contribution,
$W_V$, to the rate formula (\ref{5}) that has a substantial imaginary part in the
kinematically--allowed region. We call this term, $W_V^{(1)}$. We now take
the sum of this term and the lowest order term, $\Pi^{(0)}$, as the proper
polarization part from which to construct the ring sum, obtaining, 
\begin{equation}
\rm Im \it W_V(q,\omega)\approx\frac{\rm Im \it
W_V^{(1)}(q,\omega)}{|1-2v_1[{\rm \Pi}^{(0)}(q,\omega)+\rm Re \it W_V^{(1)}+i
\rm Im \it W_V^{(1)}(q,\omega)]|^2} \, .
\label{16}
\end {equation}
In the estimates that follow, we omit the real part of $W_V^{(1)}$ from the
denominator; it is easy to verify from (\ref{a24}) that, in contrast to the 
lowest--order imaginary part, the real part is not suppressed in the region of
$(q,\omega)$ that is important in the reaction.

Similar considerations hold for the axial--vector part. In the lepton--poor era, we
can drop the second term on the RHS of (\ref{5}), since $\omega$ will be of the order
of $T$, rather than of order 100--200 MeV, as it can be for electron neutrinos in
the trapped neutrino era \cite{tens}. Then, we define a proper $W_A$ and a contribution
$W_A^{(1)}$ as in the above and write
\begin{equation}
\rm Im\it W_A(q,\omega)\approx\frac{ \rm Im \it
W_A^{(1)}(q,\omega)}{|1-2v_2[{\rm \Pi}^{(0)}(q,\omega)+i \rm Im \it
W_A^{(1)}(q,\omega)]|^2}\, .
\label{16a}
\end {equation}

For the factors ${\rm Im} W_{V,A}^{(1)}$ we take the minimal form consistent with
avoiding a singularity in (\ref{2}) at $\omega=-\hat\mu$:
\begin{equation}
{\rm Im}W_{V,A} ^{(1)}=c^{(1)}_{V,A}(\omega+\hat\mu) \, .
\label{17}
\end{equation}
This form meets the requirement for detailed balance that  ${\rm Im}\Pi$ be odd under
the replacements, $\omega\rightarrow -\omega$, $\hat\mu\rightarrow -\hat\mu$.
 We take the parameters $c^{(1)}$ to be sufficiently small for the integrated
suppression factor to be  independent of the $[c^{(1)}]^2$ that enter through the
imaginary parts in the denominator function. In Table 2 we give, for the case of
matter at nuclear density, the separate suppression factors for the vector and
axial--vector rates generated by the uncorrected terms (\ref{17}), under
deleptonized conditions, $\mu_\nu=0$.
We note that the reductions are substantial, and conclude that the ring
corrections should be added to any model that is used for the neutrino opacities
during this era.

\section{Conclusions}

We have developed a new formalism for incorporating the effects of many--body
correlations on the charged--current rates of neutrino--matter interactions.
This formalism reveals that these rates are considerably suppressed in the densest
regions of protoneutron stars and supernova cores.  Assuming that the nucleons
are non--relativistic, our formalism incorporates the full kinematics of the 
interaction, Pauli blocking by final--state nucleons (protons), and correlation
due to nucleon--nucleon interactions.   

We have employed the ring approximation
(RPA) and assumed the near--validity of Fermi Liquid Theory.  It would desirable to
include ladder diagrams and to perform the calculations in the context of
a better numerical method for solution of the nuclear equation of state (EOS),
since the solution of the EOS is intimately related to the derivation of the
scattering/absorption rates. However, those who perform detailed nuclear EOS 
calculations and address many--body correlations in nuclear matter 
do not as yet provide the requisite
spin and density structure functions, even for the static case.   

These results for charged currents, when combined with 
the results from Burrows \& Sawyer \cite{BS} for
neutral currents, strongly suggest that energy and lepton number will leak
from supernova cores at a rate that is higher than heretofore estimated.
This implies that the neutrino luminosities during the epoch after bounce
for which the inner core is the major energy source ($> 0.5 - 1.5$s) will be 
enhanced, perhaps by as much as 50\% \cite{BS}.  The consequences of this
increased transparency for the neutrino--driven supernova explosion mechanism
\cite{bhf} may be interesting, but have yet to be clarified.   

\section*{ACKNOWLEDGMENTS}

We thank S. Reddy, M. Prakash,
G. Raffelt, and J. Lattimer for sharing their perspectives
on this class of problems and acknowledge the 
support of the NSF under grant No.
AST-96-17494.  We would also like to express our appreciation
to the Santa Barbara Institute for Theoretical Physics, supported
by the NSF under grant No. PHY94-07194.

\begin{table}
 \caption{The total suppression factors ($\cal S_{A,V}$) for the process
$\nu_e + n \rightarrow e^- + p$, for a profile in an early
post--bounce model generated by Burrows, Hayes, \& Fryxell (1995).   The 
suppression factors for the vector and the axial--vector terms are shown separately.
These suppresion factors are derived by multiplying the rate by the
neutrino occupation function and integrating over neutrino energies.
A nucleon effective mass of $0.75\times m_n$ is assumed.
}
\begin{center}
 \begin{tabular}{cccccc}
$\rho$ (g cm$^{-3}$)  & $Y_{\nu}$ & T (MeV) & $Y_e$ & $\cal S_A$& $\cal S_{V}$ \\
\hline
$3.94\times10^{14}$ & 0.077&    5  &   0.289&   0.140 & 0.269\\
$3.68\times10^{14}$ & 0.078&    5  &   0.294&   0.144 & 0.275\\
$3.08\times10^{14}$ & 0.077&    5  &   0.297&   0.157 & 0.291\\
$1.65\times10^{14}$ & 0.064&    10 &   0.275&   0.228 & 0.381\\
$2.66\times10^{13}$ & 0.01&     15 &   0.282&   0.670 & 0.775\\
$1.40\times10^{13}$ & 0.067&    15 &   0.258&   0.790 & 0.840\\
 \end{tabular}
 \end{center}
\end{table}

\begin{table}
 \caption{The total axial and vector suppression factors ($\cal S_{A,V}$) for the transformation
$\nu_e \rightarrow e^- $ during the lepton--poor era, for a density of $2.5\times10^{14}$ g cm$^{-3}$
and a neutrino chemical potential of zero.
These suppression factors are derived by multiplying the rate by the
neutrino occupation function and integrating over neutrino energies.
A nucleon effective mass of $0.75\times m_n$ is assumed.  See the text for details.
}
\begin{center}
 \begin{tabular}{cccc}
T (MeV) & $Y_e$ & $\cal S_A$& $\cal S_{V}$ \\
\hline
   3  &   0.012&   0.34 & 0.53\\
   5  &   0.013&   0.20 & 0.38\\
   7  &   0.016&   0.19 & 0.38\\
   9  &   0.018&   0.20 & 0.38\\
 \end{tabular}
 \end{center}
\end{table}

{}

\newpage

\begin{figure}
\vspace*{-1in}
\begin{center}
\leavevmode
\epsfysize=1.00\hsize
\epsfbox[160     160   612   792]{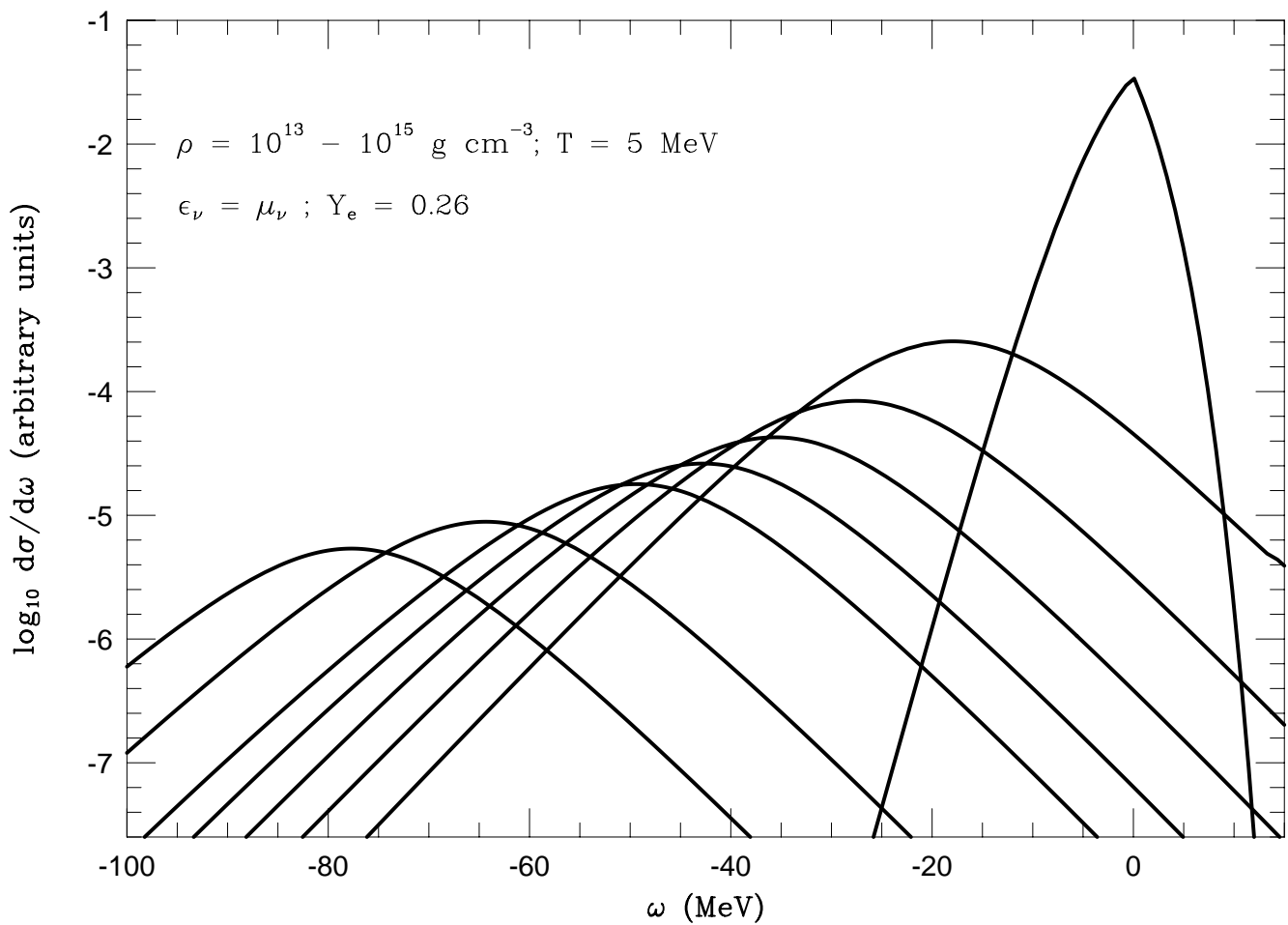}
\end{center}
\vspace*{1in}
\caption{The singly--differential cross section in arbitrary units 
versus the energy transfer, $\omega$, to the nucleons 
due to the process $\nu_e + n \rightarrow e^{-} + p$ in the
lepton--rich era.   $\omega$ is in MeV.  We have integrated Eq. (2)  
over $cos\theta$.
The curves are for mass densities of
$10^{13}$, $10^{14}$, $2\times10^{14}$, $3\times10^{14}$,
$4\times10^{14}$, $5\times10^{14}$, $7.5\times10^{14}$, and
$10^{15}$ g cm$^{-3}$.
The temperature is 5 MeV,
the electron fraction is 0.26,
the incident electron neutrinos are on their Fermi surfaces, and beta
equilibrium has been assumed.  
The $10^{13}$ g cm$^{-3}$ curve
does not include many--body effects, but does incorporate the full kinematics. 
For the other curves,  $v_2 = 4.5\times10^{-5}$ (MeV)$^{-2}$, $v_1 = 1.76\times10^{-5}$ (MeV)$^{-2}$, 
and an effective mass of $0.75\times{\rm m}_n$ was assumed.
$\hat{\mu}$ is equal to 5.3, 17.4, 26.8, 34.8, 42.1, 48.7, 63.7, 77.1 MeV for
the densities depicted.
}
\label{fig1}
\end{figure}


\begin{thebibliography}{99}
\bibitem{tens}Note that, as in  \cite{haensel}, a tensor force
 can nonetheless be the dominant ingredient in creating a proper
 polarization part that has a significant imaginary part in the region of
interest, for either the vector or axial--vector current parts.
\end{thebibliography}
\end{document}